\documentstyle[preprint,eqsecnum,aps]{revtex}
\def\test{{\raisebox{-1.0ex}{\scriptsize$r\rightarrow 0$} \atop \raisebox{0.8ex}{\scriptsize$t
\rightarrow 0$}}}
\def\be{\begin{equation}}
\def\ee{\end{equation}}
\def\beq{\begin{eqnarray}}
\def\eeq{\end{eqnarray}}
\def\n{\nonumber}
\begin{document}
\draft
\preprint{CIRI/07-swpmkg}
\title{Spherical Gravitational Collapse with Heat Flux
and Cosmic Censorship} \author{{\footnotesize S. M.
Wagh\thanks{E-mail: ciri@bom2.vsnl.net.in},\, R. V.
Saraykar\thanks{E-mail: sarayaka@nagpur.dot.net.in -- Also at
Department of Mathematics, Nagpur University, Nagpur.},\, P. S.
Muktibodh\thanks{E-mail: muktibodh@satyam.net.in -- Also at Hislop
College, Temple Road, Nagpur.}}}
\address{ Central India Research Institute, Post Box 606, Laxminagar, Nagpur
440 022, India}
\author{\footnotesize K. S.
Govinder\thanks{E-mail: govinder@nu.ac.za}} \address{ School of
Mathematical \& Statistical Sciences, University of Natal, Durban
4041, South Africa}
\date{December 10, 2001}
\maketitle

\begin{abstract}
In this paper, we investigate the nature of the singularity in the
spherically symmetrical, shear-free, gravitational collapse of a
star with heat flux using a separable metric \cite{cqg1}. For any
non-singular, regular, radial density profile for a star described
by this metric, eq. (\ref{ds2}), the singularity of the
gravitational collapse is not naked {\em locally}. Our results
here unequivocally support the Strong Cosmic Censorship
Hypothesis.

\end{abstract}

\narrowtext \clearpage
\newpage
\section{Introduction}
\label{intro} The problem of whether the gravitational collapse
leads, under general astrophysical circumstances, to a black hole
or to a naked spacetime singularity - the Cosmic Censorship
Hypothesis (CCH) - is one of the important problems of modern
physics. However, this has proven to be a very difficult question
to answer. It is therefore important for the proper mathematical
formulation of this question that we study as many of the
gravitational collapse situations as possible. It is expected that
such studies would help sharpen the statement of the CCH.

In a recent work \cite{cqg1} we have studied spherically
symmetric, separable metric spacetimes with energy/heat flux. The
source matter in those spacetimes can satisfy {\bf any} equation
of state, in particular, it is allowed to be of the form $p
=\alpha\rho$ where $p$ and $\rho$ are the pressure and the density
of the matter and $\alpha$ is a constant. We consider here the
formation of singularities in the gravitational collapse of a star
described using those spacetimes.

The organization of this paper is as follows: In \S \ref{metric}
we outline the features of the metric of the shear-free,
spherically symmetric, separable-metric spacetime with heat flux.
This class of solutions is then analyzed in \S \ref{singular} for
the existence of the strong curvature spacetime singularity. In \S
\ref{star} we argue that the solutions can model a star with heat
flow and an extended atmosphere. In \S \ref{geodesics} we analyze
the radial light cone equation for this class of metrics and show
that a horizon always forms before the singularity. We show that
for a regular, non-singular, density profile of a star, the
singularity is not even naked locally. Finally, in \S
\ref{discussion}, we discuss the implications of our results for
the Cosmic Censorship Hypothesis.

\section{The Spacetime Metric} \label{metric}
The spacetime metric of {\em all\/} the spherically symmetric,
shear-free, separable metric solutions with matter allowed to
possess the equation of state $p=\alpha\rho$ is \cite{cqg1} \be
\label{ds2} ds^2 = -\,y^2\,dt^2\;+\; R^2\left[
2\,(y^{\prime})^2\,dr^2 \;+\; y^2\,(d\theta^2
+\sin^2{(\theta)}\,d\phi^2) \right] \ee where $y\equiv y(r)$, $R
\equiv R(t)$ and an overhead prime has been used to denote the
derivative with respect to the radial coordinate $r$. The
coordinates $(t, r, \theta, \phi)$  are comoving. Note that the
spatial part of the metric (\ref{ds2}) is, in general, non-flat.

The non-vanishing components of the Ricci tensor for the metric (\ref{ds2}) are
\beq R_{00} &=& - 3\,\frac{\ddot{R}}{R} + \frac{1}{R^2} \\ \n \\
R_{01} &=& 2\,\frac{\dot{R}}{R}\frac{y^{\prime}}{y} \\ \n \\
R_{11} &=&  2\,\frac{{y^{\prime}}^2R^2}{y^2}\left( \frac{\ddot{R}}{R}
+ 2\frac{\dot{R}^2}{R^2}\right) \\ \n \\
R_{22} &=& R\dot{R} \left(\frac{\ddot{R}}{R}+2\,\frac{\dot{R}}{R}\right) \\
\n \\
R_{33} &=& \sin^2\theta R_{22} \eeq

The Ricci and the Kretschmann scalars of the metric (\ref{ds2}) are
\be {\cal R} = \frac{6}{y^2}\,\left( \frac{\ddot{R}}{R} +
\frac{\dot{R}^2}{R^2} \right) - \frac{1}{y^2R^2} \ee
\be {\cal K} \equiv R^{abcd}R_{abcd} = \frac{3 - 4\,\dot{R}^2 +
12\,{\dot{R}}^4}
   {{y}^4\,{R}^4} -
  \frac{8\,\ddot{R}}{{y}^4\,{R}^3} +
  \frac{12\,{\ddot{R}}^2}{{y}^4\,{R}^2} \ee
Clearly, the metric and the spacetime are singular for $y=0$ and/or $R=0$.

The fluid four-velocity $U^a$ and the fluid four-acceleration
$\dot{U}_a = U_{a;b}U^b$ are given by \be U^a =
\frac{1}{y}{\delta^{a}}_0 \ee \be \dot{U}_a = (0,
{{y^{\prime}}\over y},0, 0) \ee where a semicolon denotes a
covariant derivative.

We observe that the $R_{01}$ component of the Ricci tensor or,
equivalently, the $G_{01}$ component of the Einstein tensor, is
non-vanishing. This implies the presence of non-vanishing
energy-flux in the spacetime under consideration. The energy-flux
can be physically interpreted in different possible ways. Firstly,
there could be bulk motions of matter. Secondly, there could be
``thermodynamical'' heat flux arising from microphysical
considerations of matter. For the purpose of a stellar application
here, we shall consider the interpretation arising out of
thermodynamical considerations of the stellar matter.

\goodbreak
As a result, the energy-momentum tensor of the matter fluid is of
the form \be T_{ab} = (\rho + p)U_a U_b + p g_{ab} + q_a U_b
          + q_b U_a \label{u6} \ee
where $p$ is the pressure, $\rho$ is the density of the fluid and
$q^a = ( 0, q, 0, 0)$ is the radial heat flux four-vector. Note
that both shear and rotation vanish for the metric (\ref{ds2}).

The expansion, $\Theta$, heat flux, $q$, density, $\rho$ and pressure, $p$, are given by
\beq \Theta &=&\;{U^a}_{;a}\;=\; \frac{3}{y}\,\frac{\dot{R}}{R} \label{expansion}\\ \n \\
q &=& \frac{-\,1}{y^2\,y^{\prime}}\,\frac{\dot{R}}{R^3} \\ \n \\
\rho &=&\frac{1}{y^2\,R^2}\,\left[ \frac{1}{2} \;+\;3\,\dot{R}^2
\label{density} \right] \\ \n \\
p &=&
\frac{1}{2y^2R^2}\;-\;\frac{\dot{R}^2}{y^2R^2}\;-\;\frac{2\ddot{R}}{y^2R}
\label{pressure} \eeq Note that $\rho \propto 1/y^2$ and that the
radial function is not fixed by the field equations. Hence, we
can, for a non-singular density distribution, choose a nowhere
vanishing radial function $y(r)$. Note, however, that for $y(r)=r$
or $y^{\prime}=1$, the initial density profile is singular at
$r=0$ \cite{cqg1}.

From eq. (\ref{density}) and (\ref{pressure}), we obtain \be
\ddot{R}\;=\;\frac{y^2R}{6}\left[ \frac{2}{y^2R^2}\;-\;(\rho + 3p)
\right] \label{rddot} \ee It is then seen from eq. (\ref{rddot})
that the equation of state for the collapsing matter closes the
system of ordinary differential equations obtainable from the
field equations and leads to definite dynamics for the collapsing
matter.

{\em We note that, for the metric (\ref{ds2}), the equation of
state for the collapsing matter could change with the progress of
the collapse.} In other words, we emphasize that the metric
(\ref{ds2}), although initially obtained \cite{cqg1} for a
specific equation of state of the form $p=\alpha\rho$ for $\alpha$
being a constant, admits {\em any\/} equation of state for the
collapsing matter. In fact, the equation of state is an additional
input of a physical nature that is to be supplied in order to
determine the dynamics of collapsing matter.

Further, the change in the equation of state of the stellar matter
could ``halt'' the collapse and may result in a semi-stable or
stable matter configuration. However, since our interest here is
in the end-result of the unstoppable gravitational collapse, we,
in what follows, assume that the gravitational collapse is
unstoppable. Hence, our further results below apply to
``unstoppable'' gravitational collapse.

We also emphasize that in the present description we could begin
with an initial stellar density configuration that is non-singular
for all $r$. Of course, the singularity will develop as a result
of the time evolution when the physical radius of the collapsing
stellar shell vanishes at some moment of the time, {\it ie}. when
$R=0$ at some $t=t_o$, say. Further, from symmetry considerations,
such a singularity will, of course, be located at $r=0$.

\section{The Spacetime Singularity} \label{singular}
From the physical point of view, a singularity could be termed
{\em strong\/} if a test body is crushed to zero volume as it
approaches the singularity. Hence a sufficient condition for the
singularity to be a strong curvature type as defined by Tipler
{\it et al} \cite{tipler}  is: \begin{quotation} For at least one
null geodesic with an affine parameter $s$, with $s=0$ at the
singularity, the following holds \be \lim_{s \rightarrow
0}\,s^2R_{ab}K^aK^b > 0 \ee where $K^a$ is the tangent vector to
the geodesic in question and $R_{ab}$ is the Ricci tensor.
\end{quotation} If this condition is satisfied then the spacetime
is not extendible \cite{krolak}. Such spacetimes with strong
curvature singularities are to be considered in view of the CCH.

The geodesics for the metric (\ref{ds2}) are easily obtainable
\cite{kill} by considering the metric as a Lagrangian: \be 2{\cal
L}=-y^2\tilde{t}^2+R^2\left[ 2{y^{\prime}}^2\tilde{r}^2+y^2
\tilde{\theta}^2+y^2\sin^2(\theta)\, \tilde{\phi}^2\right] \ee
where an overhead tilde denotes a derivative with respect to the
affine parameter $s$ along the geodesic and $2{\cal L}$ has values
$+1, -1$ and $0$ for the spacelike, timelike and null geodesics,
respectively.

The geodesic equations are then obtainable as
\beq \frac{d}{ds}\left( Ry^2\tilde{t}\right) = -\,2{\cal L}\,\frac{dR}{dt} \\
\n \\ \frac{d}{ds}\left( R^2yy^{\prime}\tilde{r}\right) = {\cal L} \\ \n \\
\frac{d}{ds}\left[ \left( R^2y^2\tilde{\theta}\right)^2 +
k^2\cot^2{\theta} \right] = 0 \\ \n \\ \frac{d}{ds}\left(
R^2y^2\sin^2{\theta}\tilde{\phi}\right) = 0 \eeq where $k$ is a
constant of integration from the $\phi$-equation.

Let us then consider the radial null geodesics of the metric
(\ref{ds2}) for the purpose of the above condition since it
requires the existence of at least one null geodesic for its
purpose. Now, the future-directed tangent to the {\em radial
null\/} geodesic has the following components \beq
K^t \equiv \frac{dt}{ds} = \frac{K_1}{y^2 R} \label{dtds} \\ \n \\
K^r \equiv \frac{dr}{ds} = \frac{K_2}{2 y y^{\prime} R^2} \eeq
where $K_1$, $K_2$ are constants. Since $K^a$ is null, we have
${K_2}^2\,=\,2\,{K_1}^2$.

We now have \beq R_{ab}K^a K^b
\;&=&\;\left(-3\frac{\ddot{R}}{R}+\frac{1}{R^2}\right)
\frac{K_1^2}{y^4R^2}\;+\;\left(\frac{\ddot{R}}{R}
+2\frac{\dot{R}^2}{R^2}\right)\frac{K_2^2}{2y^4R^2}
\;+\;\frac{2\dot{R}K_1K_2}{y^4R^4} \\
&=&\frac{K_1^2}{y^2R^2}\,(\rho\,+\,p)\;-\;\frac{2K_1K_2}{y^2}\,y'\,R\,q
\eeq We can now, after some manipulations based on the energy
conditions, in particular, $(p+\rho)^2 > 4\,q^2$ (see Appendix),
show that \be \lim_{s\rightarrow 0}
\;s^2R_{ab}K^aK^b\;\longrightarrow \;\infty \ee Note that we have
$s=0$ at the singularity $R=0$.

Hence, the singularity of the metric (\ref{ds2}) is a strong
curvature singularity in the sense of Tipler \cite{tipler} {\it et
al}. The singularity in a spherical gravitational collapse with
heat flow considered here is then a strong curvature singularity
always.

This completes our description of the shear-free, spherically
symmetric, separable metric solutions of the Einstein field
equations that represent an inhomogeneous distribution of
imperfect matter with heat flux. Different energy conditions can
also be satisfied  for the metric (\ref{ds2}). (See Appendix
\ref{appa}. This spacetime is also of interest to cosmology
\cite{inhcos}.)

\section{Stellar Density Profile} \label{star}
Any initial density configuration of the star must be non-singular
for all $r$. For example, by noting that $\rho \propto 1/y^2$, the
radial function \cite{cqg1} could be chosen as \be y^2 = 1 +
\exp{\left(\frac{r-r_o}{L}\right)}\label{profile} \ee where $r_o$
is the boundary radius and $L$ is the thickness of the boundary
layer. The corresponding star has density that decreases with $r$
and at the boundary radius, $r_o$, there is a sharp fall in
density within the layer $L$.  Note that the density does not
vanish outside the star but can be very low as compared to that
inside the star. In the case of our sample profile, eq.
(\ref{profile}), the star has an atmosphere of exponentially
decreasing density. Essentially, the star is embedded in a
cosmological spacetime of some sort and the spacetime is not
necessarily asymptotically flat. The presence of an extended
atmosphere enveloping the star has been observed with real stars,
for example, with the Sun. {\em We emphasize that we do not assume
the profile (\ref{profile}) to obtain the results presented here.}

Also, the ``exterior'' of the star as considered here needs to be
specified. If we consider the exterior of the star as being
described by the Vaidya spacetime then we obtain the matching
condition \cite{santos} as: \begin{equation} p_{\Sigma} =
\sqrt{2}\,y\,y'|_{\Sigma} \end{equation} where $\Sigma$ is the
timelike hyper-surface across which the interior and the exterior
spacetimes are matched. This condition restricts the allowed forms
of the temporal function $R(t)$ at the stellar boundary. However,
for our purposes, it is not necessary to consider these
restrictions.

Moreover, note that $q \propto - \dot{R}/y^{\prime}$. During the
collapse, we must have $\dot{R} < 0$ (from (\ref{expansion})).
Hence, with our assumption that $y^{\prime} >0$, the heat flux is
positive, {\it ie}. heat flows from high temperature regions to
low temperature regions. Our model of a star with heat flow
therefore has a higher temperature at the center. The temperature
decreases as we move away from the stellar center.  This is what
we expect in any realistic model of a star. In our model, the
density will be a decreasing function of the radial coordinate $r$
when $y^{\prime}> 0$ and this is what we shall assume.

Of course, $y(r)$ can be constrained by other considerations. For
example, the thermal stability of the stellar structure, heat
transport equation etc.\ will, in a manner similar to that of the
Newtonian theory, yield some physical constraints on the radial
distribution of density and, hence, on $y(r)$. In fact, it is
these considerations that must be used to obtain the detailed
stellar model in a manner similar to that of the Newtonian theory.
(However, we wish to point out here that these are not relevant to
the subject of the present paper mainly because the present
considerations apply to all the forms of the function $y(r)$.)

We should also keep in mind that we require the spacetime manifold
to be locally flat in the neighborhood of the origin, $r\,=\,0$,
of the coordinate system, as it should be for any point of the
spacetime manifold. In our case, a small circle of coordinate
radius $\epsilon$ with center at the origin has circumference of
$2\pi\epsilon$. On the other hand, the circle has the proper
radius $\sqrt{2}\,y'\,\epsilon$. Then, requiring that the ratio of
the circumference to the proper radius of the circle to be $2\pi$
in the neighborhood of the origin, we obtain the condition
${y'|}_{r=0}\approx 1/\sqrt{2}$. This condition must be imposed on
$y(r)$ to obtain a realistic density profile of a star.

\section{Outgoing Radial Null Geodesics} \label{geodesics}
The radial null cone equation is \be
\frac{dt}{dr}=\pm\;\sqrt{2}\;\frac{y^{\prime}}{y}\;R \ee
where we have to choose the positive sign for the radially outgoing null geodesics.

The physical radius of the collapsing shell of matter is
$R_{Ph}=yR$. The collapsing, spherically symmetric shell of a star
must have some physical radius $R_{Ph}$ at $t=t_o$. The physical
radius becomes zero at $t=0$ signifying that the shell in question
has collapsed to a singularity. The singularity will, of course,
be at the radial coordinate $r=0$ due to symmetry considerations.
The singularity of the metric is then at $t=0$ and $r=0$.

Further, we could choose $y(r)=r$, that is, $y^{\prime}=1$.
However, as remarked earlier, this choice corresponds to a
singular initial density profile and, hence, is unacceptable to us
since a normal star does not show such a density profile.
Therefore, we use the non-singular density profile henceforth.

To determine whether the singularity is globally visible or not,
we need to look at the convergence of the principal null
congruence and determine the condition for the formation of the
outermost trapped surface. Now, the outgoing principal null vector
for the metric (\ref{ds2}) is \be \ell^a \;=\;
\frac{1}{\sqrt{2}}\,\left[ \frac{1}{y}\,{\delta^a}_t\;+\;
\frac{1}{\sqrt{2}\,y'\,R}\,{\delta^a}_r \right] \ee To test for
the existence of a trapped surface, we set the expansion rate for
$\ell^a$ to zero and look for a real solution that is positive,
$R_{ah}$. This solution corresponds to the apparent horizon - the
time history of the marginally trapped surface of the spacetime
geometry of the metric (\ref{ds2}).

Moreover, since we are dealing with separable metric functions, we
expect to obtain a condition only on the time-dependent part of
the metric, that is, $R(t)$. The condition is \be
\dot{R}\;=\;-\,\frac{1}{\sqrt{2}} \ee The trapped surface then
always covers the singularity since $\dot{R}$ must attain the
above value as it becomes (negatively) unbounded with the progress
of the collapse. Hence, the spacetime singularity is {\em not\/}
globally visible in all of the cases being considered below.

When we consider a non-singular density profile, we shall have
\[ \lim_{r \rightarrow 0}\;\frac{y^{\prime}}{y} = \ell_o \]
say, where $\ell_o$ is the appropriate positive and finite
limiting value. Now, for the geodesic tangent to be uniquely
defined and to exist at the singular point, $t=0$ and $r=0$, the
following must hold so that an outgoing, future-directed photon
trajectory exists at the singularity: \be \lim_{\test}\;\;
\frac{t}{r} = \lim_{\test}\;\; \frac{dt}{dr} = X_o \label{limits}
\ee where $X_o$ is required to be {\em real\/} and {\em positive}.
As we approach the singular point, we have, \be
X_o=\sqrt{2}\;\;\lim_{\test} \;\;\frac{y^{\prime}}{y} \,R(t)
\;=\;\sqrt{2}\;\;\lim_{ t \rightarrow 0} \;\ell_o \,R(t) \ee
Clearly, the limit in question implies that $X_o =0$. Hence, there
does not exist a real and positive tangent to the null geodesic at
the singularity if the initial density profile corresponds to a
density distribution that is non-singular.

Therefore, no null geodesics emerge from the singularity and
communicate to an external observer. Then, for any non-singular
density profile of a star with heat flux the singularity of
collapse is not visible to any observer.

\section{Discussion} \label{discussion}
A genuine spacetime singularity is a singular point of the very
structure of the spacetime. The usual laws of classical physics
break down at such a point of the spacetime manifold. It is
therefore impossible to account for the effects of the singularity
if it were to causally affect its exterior. Further, the usual
laws of quantum physics, relativistic or non-relativistic, also
fail to hold at such a spacetime location. This is mainly because
such laws have been obtained using the background spacetime arena
or the notions of space and time of the background metric. Such
notions are unavailable at the singular point of the spacetime
manifold. Unless the quantum theory of geometry or gravitation,
whatever that means, is invoked the problem of accounting for the
usual physical phenomena remains if any genuine spacetime
singularity were to causally affect its exterior. In other words,
the {\em visible\/} or {\em naked\/} spacetime singularity becomes
a genuine problem for the classical theories of physics.

It is well-known \cite{penrose1} that solutions with {\em
visible\/} or {\em naked\/} spacetime singularities can be
obtained in the General Theory of Relativity as a field theory of
gravitation. Constructing such a spacetime geometry and equating
the corresponding Einstein tensor with the energy-momentum tensor
always produces a solution of the field equations of general
relativity. However, it remains to show why such solutions cannot
be considered {\em physical\/} in some definite sense of the term,
{\it ie}.  we require some principle on the basis of which visible
spacetime singularities can be considered to be unphysical. This
is the problem of the Cosmic Censorship Hypothesis (CCH).
Moreover, since the classical theories can be considered to be
complete theories, in some appropriate sense, such a principle may
be expected to be classical in nature or content.

Thus one of the most important but as-yet unsolved problems of the
classical theory of general relativity is undoubtedly that of the
CCH. In fact, this problem is also important for the fundamental
theories of physics since it directly concerns the nature of the
gravitational field in the strong field limit. It is the same
hypothesis that is at the helm of the very existence of Black
Holes and hence Black Hole Physics, Astrophysics and
Thermodynamics as well \cite{penrose1}. As a result it is
certainly important to either prove or disprove the expectation
that this principle is of a classical nature.

Unfortunately, it has been extremely difficult to arrive at the
precise and generally agreeable formulation of the CCH let alone
prove it. This is primarily because of the conceptual nature of
the problem itself. The statement of the CCH is that {\em the
spacetime singularity be not visible to any legitimate physical
observer in a generic, realistic, physical situation such as
gravitational collapse} \cite{penrose2}. We emphasize that this is
simply an expectation based on our inability to treat a genuine
spacetime singularity in a classical theory. It necessarily
follows that if any genuine spacetime singularity were to causally
affect its exterior then the external observer would be unable to
account for the physical phenomena in its vicinity. The classical
predictability would then be in jeopardy if the singularity were
{\em visible\/} to the external observer.

As regards the statement of the hypothesis, the real issue then is
about making the terms such as {\em generic\/}, {\em physical\/}
or {\em realistic\/} precise so as to make some mathematically
provable or disprovable statement for this principle of Cosmic
Censorship. For example, we could demand that the collapsing
matter satisfy some energy conditions/equation of state, that the
spacetime not possess any special symmetry, that it should be
stable to perturbations and so on. However, it has proven to be
rather difficult to translate such demands into any generally
agreeable mathematically precise statements.

In the absence of any precise formulation of the CCH and hence its
proof, we can, at  best, look for examples of naked or visible
singularities. From such examples, we can hope to sharpen the
meaning of various terms in the statement of the CCH.

In the recent past, many workers \cite{psj1} have followed this
approach. However, we note that none of these examples can be
considered to be a genuine counter-example to the CCH since all
can be considered to be {\em special\/} in some or the other
sense. Nonetheless, it is true that such examples can be used to
provide some definite meaning to the statement of the censorship
hypothesis.

To illustrate the above point, we note that the locally naked
singularity of the Vaidya-de Sitter metric can be shown
\cite{wagh1} to be the same as that of the Vaidya metric. Then, we
can argue that the asymptotic flatness of the solution does not
manifest, in some sense, in the occurrence or not of the naked
spacetime singularity, {\it ie}.  the asymptotic observer has no
role to play in the formulation of the censorship hypothesis.
Therefore, this example unequivocally supports the formulation of
the strong CCH due to Penrose \cite{penrose1}. Essentially, the
Strong CCH demands that the space-time singularity be not visible
to any observer.

In this paper, we considered the shear-free collapse of a
spherical star with heat flux using the space-time of the metric
(\ref{ds2}). Moreover, our model, based on the metric (\ref{ds2}),
can accommodate the realistic feature that the star can possess an
extended atmosphere similar to what is observed with the real
stars. Our results show that the singularity resulting in the
stellar collapse modelled using the metric (\ref{ds2}) is not
visible to {\em any observer\/} if the initial density profile of
the star is non-singular similar to what is observed with real
stars. The strong cosmic censorship hypothesis states that the
singularity be not visible to any observer. Hence, our results are
in complete accord with the Strong CCH for the non-singular
density profile of the stellar model considered here.

\acknowledgments We are grateful to all the participants of the
Discussion Conference on Gravitational Collapse held by CIRI at
Nagpur during March 7-10, 2000 for many stimulating and marathon
discussions. One of us, SMW, also thanks Ramesh Narayan and Pankaj
Joshi for discussion on the issue of naked singularities during
the ICGC-2000 at Kharagpur, India. He  also   thanks Jayant
Narlikar and Naresh Dadhich for discussions that clarified some of
the ideas presented here. KSG thanks the University of Natal for
ongoing support.

\newpage
\appendix
\section{} \label{appa}
In this appendix, we discuss the energy conditions as obtainable for the
considered spacetime. (For an elegant analysis and further details of
energy conditions applicable to imperfect fluids, see \cite{kst}.)

We recall the definitions of the energy conditions for the sake of
completeness:
\begin{itemize}
\item {\em Weak energy condition}: $T_{ab}W^aW^b \ge 0$ where $W^a$ is a
timelike, future-directed 4-vector. It implies that the energy density as
measured by any observer is {\em positive}.
\item {\em Dominant energy condition}: $F_a\;=\;T_{ab}W^b$ must be
future-directed, timelike or null for any timelike, future-directed $W^a$.
It implies that the speed of energy flow of matter must be less than the
speed of light for every observer.
\item {\em Strong energy condition}: $2\,T_{ab}W^aW^b\;+\;T\;\ge\;0$ for
any timelike, future-directed, unit 4-vector $W^a$ where $T$ is the trace
of $T_{ab}$. \end{itemize}
We may note here that the Dominant energy condition implies the Weak energy
condition, that the Strong energy condition can be violated in the
presence of negative pressure, that is, when $T_{ab}W^aW^b$ is negative
and that the investigation of energy conditions is an algebraic problem
that leads to a search of the roots of a polynomial of degree 4 in a four
dimensional spacetime.

From \cite{kst}, we also recall here the corollary for the shear-free or
dynamic-viscosity-free, spherically symmetric spacetimes. We note that
the metric (\ref{ds2}) describes a shear-free spacetime.

\goodbreak
Fluid that undergoes shear-free motion or has vanishing coefficient of
dynamic viscosity satisfies \begin{itemize}
\item the Weak energy condition iff
\be \rho + p \ge 2q  \label{a1} \ee
\be \rho - p + \triangle > 0 \ee
\item the Dominant energy condition iff, in addition to (\ref{a1}),
we have \be \rho - p \ge 0 \;\;\;\;\;\;\;\;\;\;\;\;\;\;\rho - 3p +
\triangle \ge 0 \ee
\item the Strong energy condition iff, in addition to (\ref{a1}),
we have \be 2p + \triangle \ge 0\ee \end{itemize}
where $\triangle \,=\,\sqrt{(p+\rho)^2\,-\,4q^2}$.

In general, these conditions are to be imposed on the spacetime
under consideration. However, it is easy to verify that these
conditions are satisfied for the spacetime of the metric
(\ref{ds2}) when $0 \le \alpha \le 1$ for the simple equation of
state $p\;=\;\alpha\rho$.

\bigskip
\bigskip

\noindent (Submitted to: General Relativity \& Gravitation))
\end{document}